\documentclass[a4paper,showpacs]{iopart}
\pdfoutput=1 
\usepackage[utf8]{inputenc}
\usepackage[T1]{fontenc}
\usepackage[usenames,dvipsnames]{xcolor}
\usepackage{graphicx}
\usepackage{iopams}
\usepackage{amsbsy}

\usepackage{color}
\definecolor{darkblue}{rgb}{0,0,0.6}
\definecolor{darkred}{rgb}{0.6,0,0}

\usepackage[colorlinks=true,urlcolor=darkblue,citecolor=darkblue,linkcolor=darkred,hyperfootnotes=false]{hyperref}

\newcommand{\ind}[1]{_{\mathrm{#1}}}

\newcommand{\sinc}{{\rm \,sinc\!}}

\newcommand{\dd}{\mathrm{d}}
\newcommand{\ed}{\mathrm{e}}

\renewcommand{\emph}{\textit}
\newcommand{\binom}[2]{\left( {\begin{array}{c} \!\!\!{#1}\!\!\! \\ \!\!\!{#2}\!\!\! \end{array}} \right) }

\begin{document}

\title{Effect of disorder geometry on the critical force in disordered elastic systems}


\author{Vincent D\' emery$^{1,2}$}
\address{$^1$ Institut Jean Le Rond d'Alembert (UMR CNRS 7190), Universit\'e Pierre et Marie Curie, F-75005 Paris, France}
\address{$^2$ Department of Physics, University of Massachusetts, Amherst, MA 01003, USA.}
\ead{vdemery@physics.umass.edu}

\author{Vivien Lecomte}
\address{Laboratoire Probabilités et Modèles Aléatoires (UMR CNRS 7599), Université Pierre et Marie Curie \& Université Paris Diderot, 75013 Paris, France}

\author{Alberto Rosso}
\address{Laboratoire Physique Th{\'e}orique et Mod{\`e}les  Statistiques (UMR CNRS 8626), Universit\'e de Paris-Sud,
Orsay Cedex, France}

\begin{abstract}
We address the effect of disorder geometry on the critical force in disordered elastic systems. We focus on the model system of a long-range elastic line driven in a random landscape. 
In the collective pinning regime, we compute the critical force perturbatively.
Not only our expression for the critical force confirms previous results on its scaling with respect to the microscopic disorder parameters, it also provides its precise dependence on the disorder geometry (represented by the disorder two-point correlation function).
Our results are successfully compared to the results of numerical simulations for random field and random bond disorders. 
\end{abstract}


\maketitle

\section{Introduction}

\begin{figure}
\begin{center}
\includegraphics[width=.650\columnwidth]{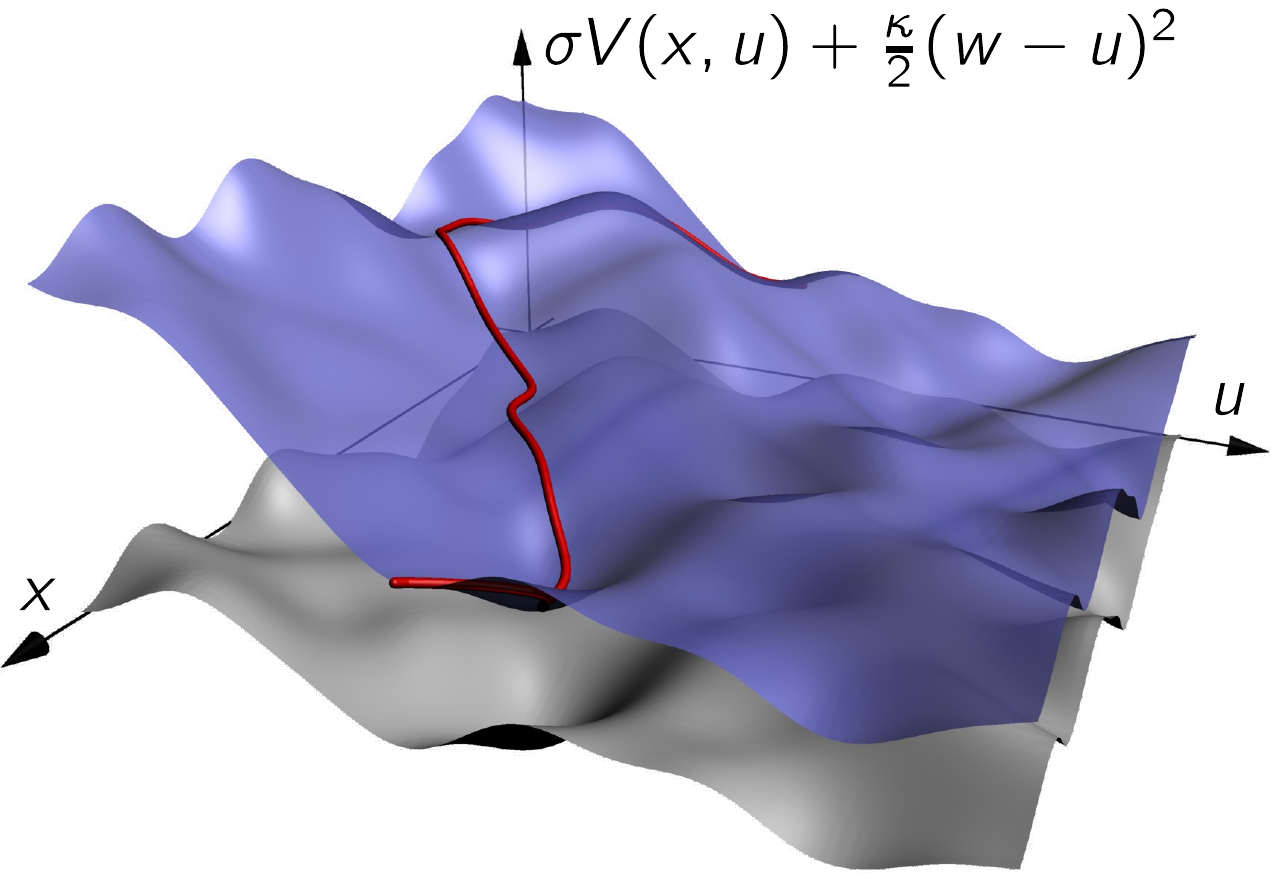}
\end{center}
\caption{(Colour online) An elastic line pulled by a spring of stiffness~$\kappa$ and position~$w$ in a random landscape. The bottom grey surface is the potential $\sigma V(x,u)$. The top blue surface is the effective potential seen by the line, \textit{i.e.}~the bare potential plus the parabolic potential $\frac{\kappa}{2}(w-u)^2$ exerted to the spring.}
\label{fig_schema}
\end{figure}

Disordered elastic systems~\cite{halpin-healy_kinetic_1995, Kardar1998, brazovskii_pinning_2004, Agoritsas2012} are ubiquitous in Nature and condensed matter physics; they encompass a wide range of systems going from vortex lattices in superconductors~\cite{Larkin1979} to ferromagnetic domain walls~\cite{Lemerle1998}, wetting fronts~\cite{Joanny1984}, imbibition fronts~\cite{Soriano2002, Santucci2011} or crack fronts in brittle solids~\cite{Bouchaud2002, Bonamy2008}. In simple models for those phenomena, an elastic object struggles to stay flat while its random environment tries to deform it, in or out of equilibrium. 
An example is given by an elastic line in a random landscape, that is pictured on Fig.~\ref{fig_schema}. 
As a result of the competition between disorder and elasticity, the elastic object becomes rough and is characterized by a universal roughness exponent~\cite{barabasi_fractal_1995,krug_origins_1997} that depends on the dimension of the problem, the range of the elastic interaction and the type of disorder, but not on the microscopic details of the system~\cite{Kardar1998, Agoritsas2012}.

This coupling between disorder and elasticity also has an important consequence on the response of the elastic object to an external force~\cite{Larkin1979}. At zero temperature, there exists a critical force below which it does not move and remains \textit{pinned} by the disorder. If the applied force is larger than this threshold, the elastic object \textit{unpins} and acquires a non-zero average velocity. This describes the \textit{depinning transition} of the elastic line. A finite temperature rounds this behaviour for forces close to the threshold~\cite{Bustingorry2008} and allows the object to move at a finite velocity for forces well below the threshold, by a thermally activated motion called \textit{creep}~\cite{Ioffe1987,nattermann_scaling_1990,balents_large-n_1993,Chauve2000,Kolton2009}.
The critical force plays a crucial role in applications.
In {type-II} superconductors, it corresponds to the critical current above which the vortex lattice starts moving, leading to a superconductivity breakdown~\cite{Anderson1964}. In brittle solids, it determines the critical loading needed for a crack to propagate through the sample and break it apart~\cite{Bouchaud2002}.

Contrarily to roughness and depinning exponents, the critical force is not a universal quantity and its value depends in general on the details of the model. 
However, scaling arguments allow to find its dependence on the disorder amplitude and the different lengthscales present in the system, such as the size of the defects and the typical distance between them~\cite{Larkin1979, Nattermann1990}.
Unfortunately this approach gives the critical force only up to a numerical prefactor, whose value depends on microscopic quantities such as the geometrical shape of the impurities, and which is essential to determine in view of applications. Recently, a numerical self-consistent  scheme~\cite{Roux2003,Roux2008} and numerical simulations have focused on a precise determination of the critical force~\cite{Demery2012c,Patinet2013} in the context of brittle failure.
Notably, is has been shown that in the \textit{collective} regime, occurring at weak disorder amplitude, the critical force does not depend on the disorder distribution but only on the disorder amplitude and correlation length~\cite{Demery2012c}. Still, the effect of the disorder geometry, that is partly encoded in its two-point correlation function, remains to be determined.

In this article, we address the question of the dependence of the critical force on the disorder geometry. We focus on the case of a long-range elastic line in a random potential, that is the relevant model for wetting fronts and crack fronts in brittle failure. 
We restrict ourselves to the collective pinning regime, that appears when the disorder amplitude is small. 
The line is driven by a spring pulled at constant velocity, and the \textit{drag force} needed to move the spring is computed perturbatively in the disorder amplitude. In the limit of zero spring stiffness and zero velocity, this force is the critical force and we derive its analytic expression.

Our expression depends explicitly on the two-point correlation function of the disorder, and thus on the disorder geometry. Moreover, our computation is valid for a random bond disorder as well as a random field disorder~\cite{Chauve2000}. Numerical simulations are performed for both types of disorder and various disorder geometries. They provide a successful check of our analytical result and show that two systems with the same disorder amplitude and correlation length can have a different critical force if their two-point correlation functions are different.

The paper is organized as follows. In section~\ref{sec_model}, we introduce the model of a long-range elastic line driven in a random landscape. In section~\ref{sec_main_result}, we summarize our results. Section~\ref{sec_analytical} is devoted to the analytical computation of the drag force, from which we deduce the critical force. Numerical simulations details and results are presented in section~\ref{sec_numerical}. We conclude in section~\ref{sec_conclu}.

\section{Model}\label{sec_model}

We consider a 1+1 dimensional elastic line of internal coordinate $x$ and position $u(x,t)$, pulled by a spring of stiffness $\kappa$ located at position $w(t)$ in a random energy landscape $\sigma V(x,u)$; this system is represented on Fig.~\ref{fig_schema}. The parameter $\sigma$ represents the disorder amplitude. 
The equation of evolution of the line position at zero temperature is~\cite{Kardar1998, Ferrero2013} 
\begin{equation}\label{eq_evol}
\partial_t u(x,t)=\kappa[w(t)-u(x,t)]+f\ind{el}[u(\cdot,t)](x)-\sigma \partial_u V(x, u(x,t)).
\end{equation}
The elastic force $f\ind{el}[u(\cdot,t)](x)$ is linear in $u(x,t)$ and we consider the case of a long-range elasticity
\begin{equation}\label{eq_force_el_cont}
f\ind{el}[u(\cdot,t)](x)=\frac{c}{\pi}\int \frac{u(x',t)-u(x,t)}{(x-x')^2}\dd x'.
\end{equation}
The disorder has zero mean ($\overline{V(x,u)}=0$) and two-point correlation function
\begin{equation}
\overline{V(x,u)V(x',u')} =R_x(x-x')R_u(u-u').
\end{equation}
The overline represents the average over the disorder.
Alternatively, one can also use the force correlation function, 
\begin{eqnarray}
\overline{\partial_u V(x,u)\partial_{u'}V(x',u')} & =- R_x(x-x')\partial_u^2 R_u(u-u')
\label{eq:2pointforcecorrelator}\\
&
 =\Delta_x(x-x')\Delta_u(u-u').
\label{eq:2pointforcecorrelatorDelta}
\end{eqnarray}
with $\Delta_u=-\partial_u^2 R_u$.
We do not assume that the disorder is Gaussian distributed.
For the so-called random bond (RB) case, the potential $V(x,u)$  is short-range correlated both in the variables $x$ and $u$; this implies a global constraint on the force correlation function, namely  $\int \Delta_u(u) \dd u=0$. 
For the random field (RF) case, $V(x,u)$ is for instance a Brownian motion 
as a function of $u$, with diffusion constant   $\int \Delta_u(u) \dd u>0$~\cite{Chauve2000}.

Finally, we impose a constant velocity $v$ to the spring,
\begin{equation}
w(t)=vt.
\end{equation}
The drag $f\ind{dr}$ is defined to be the average force exerted on the line by the spring
\begin{equation}
f\ind{dr}(\kappa,v)=\kappa\left[w(t)- \overline{\left\langle u(x,t) \right\rangle}\right],
\label{eq:def_fdrag}
\end{equation}
where $\langle \cdot \rangle$ denotes the average along the internal coordinate $x$. Since the landscape is statistically translation invariant, this quantity is expected not to depend on time.
Besides, it is expected that both the drag force and the critical force, which depend on the realization of the disorder, tend to a limit at large system size which is independent of the particular realization --~hence equal to its average over disorder, as we make use of in~(\ref{eq:def_fdrag}).
Those averages are the so-called \emph{thermodynamic} drag and critical forces.

We focus on the computation of the drag as a function of the spring stiffness $\kappa$ and velocity $v$. We then show how to extract the critical force from this force-velocity characteristics, fixing the velocity instead of the force.

\section{Main result}\label{sec_main_result}

\begin{figure}
\begin{center}
\includegraphics[width=.650\columnwidth]{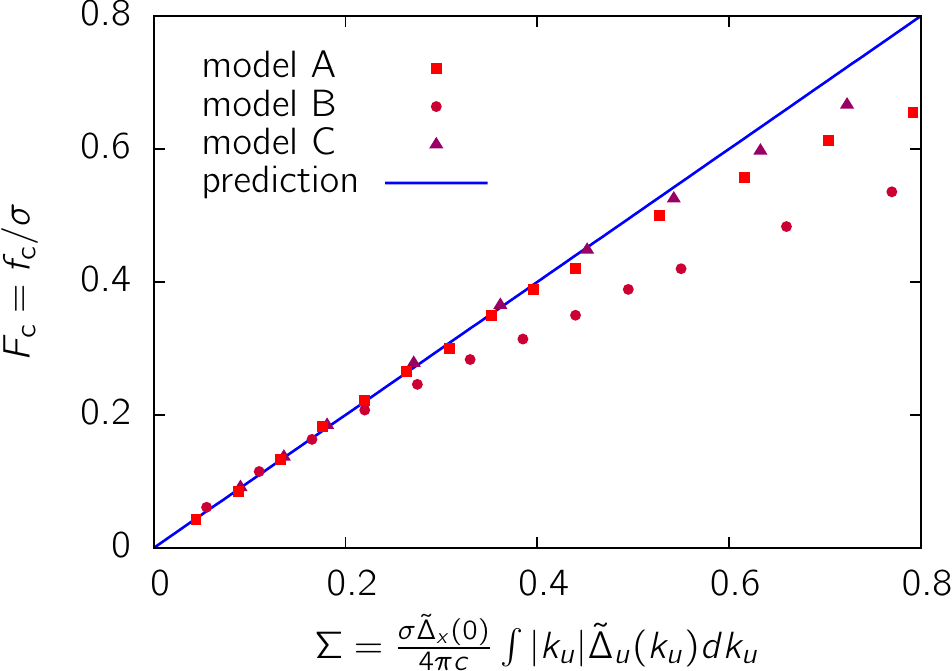}
\end{center}
\caption{(Colour online) Critical force as a function of the dimensionless disorder amplitude $\Sigma$, defined in~(\ref{eq:disorder_parameter}): comparison between numerical simulations and analytical prediction (\ref{eq_anal_pred}) for random field (models A and B) and random bond (model C) disorders (the disorders are defined precisely in section \ref{sub_numerical_model}). Points are results of the simulations and the line is the analytical prediction, valid in the small $\Sigma$ limit.}
\label{fig_sigma_fc}
\end{figure}

Our main result is the following expression of the critical force, valid in the collective pinning regime (when the disorder amplitude $\sigma$ is small):
\begin{equation}\label{eq_anal_pred}
f\ind{c} \simeq \frac{\sigma^2\tilde \Delta_x(0)}{4\pi c}\int|k_u|\tilde \Delta_u(k_u)\dd k_u.
\end{equation}
where $\tilde \Delta_{x,u}$ are Fourier transforms of $\Delta_{x,u}$.
The expression of~$f\ind{c}$ holds for a long-range elastic line for both random bond and random field disorders.
It is compared to simulations results on Fig.~\ref{fig_sigma_fc}, that shows a very good agreement in the collective pinning regime~$\Sigma\ll 1$.
The dimensionless disorder amplitude $\Sigma$ is defined later (\ref{eq:disorder_parameter}).

This analytic result is derived in Section~\ref{sec_analytical} and the numerical simulations are detailed in section~\ref{sec_numerical}.

\section{Analytical computation}\label{sec_analytical}

In this section, we start by computing the average force required to drive the line at an average speed $v$ with a spring of stiffness $\kappa$. This provides us a force-velocity characteristic curve that depends on the spring stiffness. 
From these characteristic curves, we deduce that sending the velocity and stiffness to zero in the appropriate order allows to extract the critical force.

\subsection{Drag force}

The line evolution equation (\ref{eq_evol}) is highly non linear due to the presence of the random potential; it is thus very difficult to handle. To evaluate the average drag, we resort to a perturbative analysis in the disorder amplitude~$\sigma$.

We expand the line position in powers of $\sigma$ as
\begin{equation}
u(t)=\sum_n \sigma^n u_n(t).
\end{equation}
At order $0$, the solution of~(\ref{eq_evol}) is independent of disorder
\begin{equation}\label{eq_zeroth_order}
u_0(x,t)=vt-\frac{v}{\kappa},
\end{equation}
leading from~(\ref{eq:def_fdrag}) to the average drag
\begin{equation}\label{eq_drag_0}
f\ind{dr}^{(0)}(\kappa,v)=v.
\end{equation}

The computation at higher orders is done in Fourier space, with the convention $\tilde g(k_x)=\int g(x) \ed^{-ik_xx} \dd x$ (and similarly along direction $u$). We start by Fourier transforming the elastic force,
\begin{equation}
f\ind{el}[u(\cdot,t)](x)=-c\int |k_x|\tilde u(k_x,t)\ed^{ik_xx}\frac{\dd k_x}{2\pi},
\end{equation}
and the disorder correlator,
\begin{equation}\label{eq_dis_correl_fourier}
\overline{\tilde V(k_x,k_u)\tilde V(k_x',k_u')} =\\(2\pi)^2\delta(k_x+k_x')\delta(k_u+k_u')\tilde R_x(k_x)\tilde R_u(k_u).
\end{equation}
We also define, corresponding to~(\ref{eq:2pointforcecorrelator}-\ref{eq:2pointforcecorrelatorDelta})
\begin{eqnarray}
\tilde \Delta_x(k_x) & = \tilde R_x(k_x),\\
\tilde \Delta_u(k_u) & = k_u^2\tilde R_u(k_u).
\end{eqnarray}

The first order contribution to the line position satisfies
\begin{equation}
\partial_t u_1(x,t)+\kappa u_1(x,t)-f\ind{el}[u_1(\cdot,t)](x)=-\partial_u V(x,u_0(t)).
\end{equation}
Fourier transforming this equation in directions $x$ and $u$ gives the solution in Fourier space
\begin{equation}\label{eq_line_first_order}
\tilde u_1(k_x,t)=-\int \frac{ik_u}{ik_u v+\omega(k_x)}\ed^{ik_u u_0(t)}\tilde V(k_x,k_u)\frac{\dd k_u}{2\pi},
\end{equation}
where we have introduced the damping rate
\begin{equation}\label{eq_damping_cont}
\omega(k_x)=\kappa+c|k_x|,
\end{equation}
which fully encompasses the effect of the elasticity.
Since the first order correction is linear in the potential~$V$, its average over disorder is 0 and it does not contribute to the drag: $f\ind{dr}^{(1)}=0$.

At second order, the evolution equation reads
\begin{eqnarray}
\partial_t u_2(x,t)+&\kappa u_2(x,t)-f\ind{el}[u_2(\cdot,t)](x) =\nonumber\\
  &-\sigma^{-1} [\partial_u V(x,u_0(t)+\sigma u_1(x,t))-\partial_u V(x,u_0(t))].
\end{eqnarray}
Following an idea introduced by Larkin~\cite{Larkin1970}, we expand the potential around $u_0(t)$, getting
\begin{eqnarray}
\partial_t u_2(x,t)+\kappa u_2(x,t)-f\ind{el}[u_2(\cdot,t)](x)=-\partial_u^2 V(x,u_0(t))u_1(x,t).
\label{eq_expansion}
\end{eqnarray}
It reads in Fourier space
\begin{eqnarray}
\partial_t\tilde u_2(k_x,t) + &\omega(k_x)\tilde u_2(k_x,t)=\nonumber\\ &
\int k_u^2 \ed^{ik_u u_0(t)}\tilde V(k_x',k_u)\tilde u_1(k_x-k_x',t)\frac{\dd k_x' \dd k_u}{(2\pi)^2}.
\end{eqnarray}
Inserting the first order result (\ref{eq_line_first_order}) and solving gives
\begin{eqnarray}
\tilde u_2(k_x,t) =& \int \frac{ik_u'^2(k_u'-k_u) \ed^{ik_u u_0(t)}}{[ik_uv+\omega(k_x)][i(k_u-k_u')v+\omega(k_x-k_x')]} \nonumber\\ &
\ \qquad \times \tilde V(k_x-k_x',k_u-k_u')\tilde V(k_x',k_u')\frac{\dd k_x'\dd k_u\dd k_u'}{(2\pi)^3}.
\end{eqnarray}
Averaging over disorder with (\ref{eq_dis_correl_fourier}) leads to
\begin{equation}
\overline{\tilde u_2(k_x)}=2\pi\delta(k_x)\int \frac{ik_u\tilde \Delta_x(k_x')\tilde \Delta_u(k_u)}{\kappa[-ik_uv+\omega(k_x')]} \frac{\dd k_x'\dd k_u}{(2\pi)^2}.
\end{equation}
The proportionality to $\delta(k_x)$ signifies that, in direct space, the average second order correction does not depend on the internal coordinate $x$. It reads
\begin{equation}
\overline{u_2}=-\int \frac{vk_u^2\tilde \Delta_x(k_x)\tilde \Delta_u(k_u)}{\kappa[k_u^2v^2+\omega(k_x)^2]} \frac{\dd k_x \dd k_u}{(2\pi)^2}.
\end{equation}
The second order drag is thus:
\begin{equation}
f\ind{dr}^{(2)}(v,\kappa)=\sigma^2\int \frac{vk_u^2\tilde \Delta_x(k_x)\tilde \Delta_u(k_u)}{k_u^2v^2+(\kappa+c|k_x|)^2} \frac{\dd k_x \dd k_u}{(2\pi)^2}.
\end{equation}
Adding this result to the zeroth order drag (\ref{eq_drag_0}) provides the drag up to the order $\sigma^2$,
\begin{eqnarray}
f\ind{dr,2}(v,\kappa)& = f\ind{dr}^{(0)}(v,\kappa) + f\ind{dr}^{(1)}(v,\kappa) + f\ind{dr}^{(2)}(v,\kappa) \\
& = v+\sigma^2\int \frac{vk_u^2\tilde \Delta_x(k_x)\tilde \Delta_u(k_u)}{k_u^2v^2+(\kappa+c|k_x|)^2} \frac{\dd k_x \dd k_u}{(2\pi)^2}. \label{eq_drag_line}
\end{eqnarray}

\begin{figure}
\begin{center}
\includegraphics[width=.650\columnwidth]{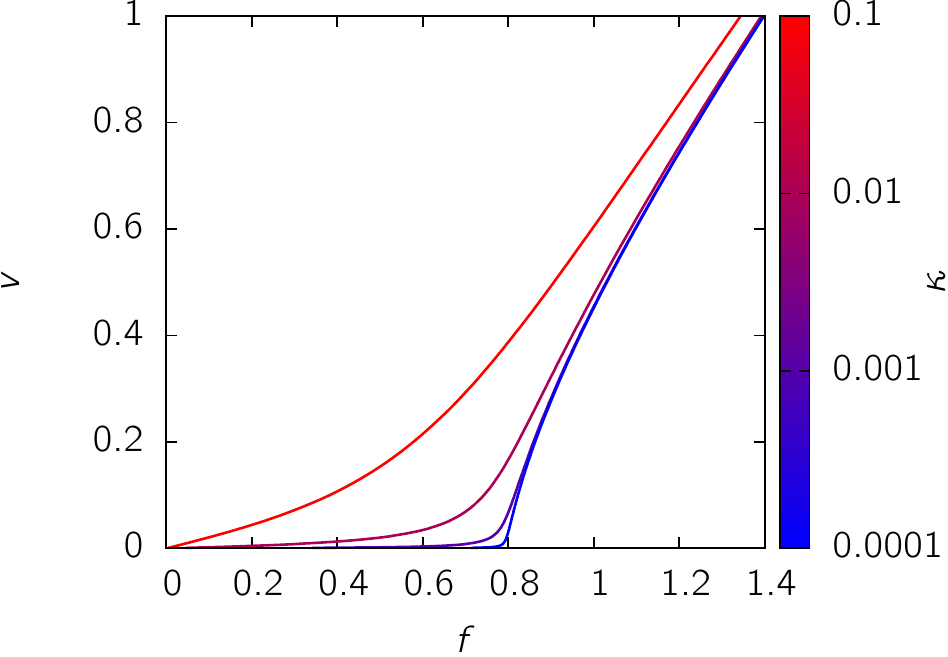}
\end{center}
\caption{(Colour online) Force-velocity curve using the drag~(\ref{eq_drag_line}) computed to the second order in $\sigma$, for different values of the parabola curvature $\kappa$, indicated by the colour scale. The two-point functions $\Delta_x$ and $\Delta_u$ of the disorder are centered Gaussian functions of unit variance, encoding a random field disorder.}
\label{fig_f_v}
\end{figure}

This drag gives us access, at the perturbative level, to the crucial force-velocity characteristic. It is plotted on Fig.~\ref{fig_f_v} for a  random field disorder with Gaussian two-point functions, at different values of the spring stiffness. 
When the spring stiffness $\kappa$ goes to zero, the depinning transition appears clearly and becomes sharp when $\kappa=0$. The picture is qualitatively similar for a random bond disorder. Any positive stiffness rounds the transition, analogously to the temperature~\cite{Agoritsas2012, Chen1995, Bustingorry2008}.

\subsection{Critical force}\label{label}

As noted above, the usual force-velocity characteristic at zero temperature is recovered in the limit $\kappa\rightarrow 0$; its equation is given by
\begin{equation}\label{eq_v_f_k0}
f_{\kappa=0,2}(v)=v+\frac{\sigma^2}{c}\int \frac{k_u^2\tilde\Delta_x(vq/c)\tilde\Delta_u(k_u)}{k_u^2+q^2}\frac{\dd q\dd k_u}{(2\pi)^2}.
\end{equation}
We have performed the variable substitution $ck_x=vq$ in order to eliminate the velocity in the denominator.
Taking the small velocity limit in this expression gives the critical force (\ref{eq_anal_pred})
\begin{equation}
f\ind{c,2}=\frac{\sigma^2\tilde \Delta_x(0)}{4\pi c}\int|k_u|\tilde \Delta_u(k_u)\dd k_u.
\end{equation}
This expression is our main result, announced in Eq.~(\ref{eq_anal_pred}).
The index $2$ indicates that this critical force comes from a second order perturbative computation in the disorder amplitude $\sigma$. The two limits do not commute: since any non-zero stiffness rounds the transition, taking the limit of zero velocity first would give a zero critical force (see Fig.~\ref{fig_f_v}).
To get the depinning exponent $\beta$ defined by $v \sim_{f\rightarrow f\ind{c}^+} (f-f\ind{c})^\beta$, we have to go one step further in the Taylor expansion of (\ref{eq_v_f_k0}) around $v=0$. This can be done analytically for simple correlators $\tilde\Delta_x(k_x)$ and $\tilde\Delta_u(k_u)$, or numerically in the general case (see Fig.~\ref{fig_f_v} for an example). We get
\begin{equation}
v\sim f-f\ind{c},
\end{equation}
which corresponds to the mean field behaviour ${\beta=\beta\ind{MF}=1}$~\cite{Fisher1998} valid above the upper critical dimension $d\ind{uc}$ (for the long-range elasticity $d\ind{uc}=2$). Below $d\ind{uc}$,  the mean field value $\beta\ind{MF}=1$ is an upper bound of the exact value of $\beta$ which can be estimated by a functional renormalization group $\epsilon-$expansion~\cite{Chauve2001,le_doussal_functional_2004,wiese_functional_2006} or evaluated numerically to $\beta=0.625 \pm 0.0005$ for the long-range elastic line~\cite{Duemmer2007}.

\section{Numerical simulations}\label{sec_numerical}

We now turn to the comparison of our analytical prediction to numerical simulations of the line. 
Since our computation remains valid for both  random bond and random field  disorder, we perform numerical simulations on both cases.

\subsection{Numerical model}\label{sub_numerical_model}

In our model a line of length $L$ is discretized with a step $a$ and its elasticity is given by
\begin{equation}\label{eq_force_el_disc}
f\ind{el}[u(\cdot,t)]_n=\frac{c}{\pi a}\sum_{n'\neq 0} \frac{u_{n'}(t)-u_n(t)}{(n-n')^2}.
\end{equation}
Each point of the line moves on a rail with a disordered potential which is uncorrelated with the others rails, so that $\Delta_x(x-x') = \delta_{x,x'}$. Three different models of disorder are considered: 
\begin{itemize}\setlength{\itemsep}{1pt}
\item model A: random field disorder obtained by the linear interpolation of the random force drawn at the extremities of  segments of length $1$; 
\item model B: random field disorder obtained in the same way as model A, but the segments have length $0.1$ with probability $1/2$ and $1.9$ with probability $1/2$;
\item model C: random bond disorder obtained by the spline interpolation of the random energies drawn at the extremities of  segments of length $1$. 
\end{itemize}

\subsection{Analytical prediction}

The prediction of the critical force for the three models can be obtained  observing that  for a discrete line the damping rate (\ref{eq_damping_cont}) changes to
\begin{equation}
\omega(k_x)=\kappa+c \left(|k_x|- \frac{ak_x^2}{2\pi}\right).
\end{equation}
where the wave-vector  $k_x$ is restricted to $\left[-\pi/a,\pi/a \right]$.
The limits $\kappa\rightarrow 0$ and $v\rightarrow 0$ give exactly the same result as (\ref{eq_anal_pred}):
\begin{equation}
f\ind{c,2}=\lim_{v\rightarrow 0}\lim_{\kappa\rightarrow 0} f\ind{tot,2}(v,\kappa)=\frac{\sigma^2\tilde \Delta_x(0)}{4\pi c}\int|k_u|\tilde \Delta_u(k_u)\dd k_u.
\end{equation}
It is remarkable that discretizing the line does not change the critical force. In all models we set $a=1$ so that $\tilde\Delta_x(0)=1$ and the functional of $\Delta_u(u)$ appearing in our expression (\ref{eq_anal_pred}) for the critical force is computed in \ref{ap_disorder_correlations} for the three models. The final prediction for model A is:
\begin{equation}\label{eq_anal_pred_044}
f\ind{c,2}=\frac{2\log(2)}{\pi}\frac{\sigma^2}{c}\simeq 0.44\frac{\sigma^2}{c}.
\end{equation}
while for model B a numerical computation gives
\begin{equation}
f\ind{c,2}\simeq 0.55\frac{\sigma^2}{c},
\end{equation}
and for model C we have
\begin{equation}
f\ind{c,2}\simeq 2.83 \frac{\sigma^2}{c}.
\end{equation}

\subsection{Measurement of the critical force}

\begin{figure}
\begin{center}
\includegraphics[width=.650\columnwidth]{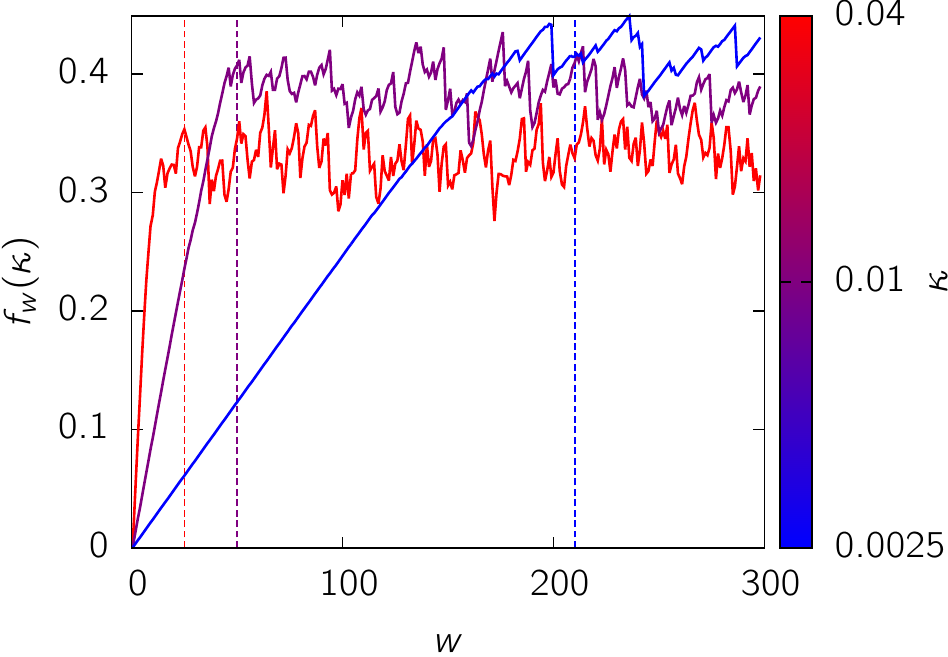}
\end{center}
\caption{(Colour online) Pinning force (\ref{eq_def_pin_force}) as a function of the position~$w$ of the parabola, for different values of the spring stiffness $\kappa$. The dashed vertical lines represent approximate values $w^*$ after which the pinning force becomes independent of the initial condition. Model A disorder with $\sigma=1$ is used here.}
\label{fig_w_f}
\end{figure}

We start our numerical procedure with a flat configuration $u(x)=0$ and $w=0$. 
Then the interface moves to a state,
$u_{w=0}(x)$ which is stable with respect to small deformations. Increasing  $w$, the interface position increases and a sequence of stable states can be recorded.
For each $w$, the stable state $u_w(x)$ can be found using the algorithm proposed in~\cite{Rosso2002} and we measure the pinning force 
\begin{equation}\label{eq_def_pin_force}
f_w(\kappa)=\kappa [w-\langle u_w(x)\rangle].
\end{equation}
This pinning force depends on the realization of the disordered potential (its fluctuations have been studied in~\cite{bolech_universal_2004}).
An example of the evolution of the pinning force with $w$ is shown on Fig.~\ref{fig_w_f}.
The pinning force is in general dependent on the initial condition, however, due to the Middleton no-passing rule~\cite{Middleton1992}, we can prove that there exists a $w^*>0$ such that the sequence of stable states $u_{w>w^*}(x)$ becomes independent of the initial condition. 
A stationary state is thus reached, where the pinning force oscillates around its average value $\overline{f(\kappa)}$ and displays correlation in $w$. 
Thus in order to estimate correctly $\overline{f(\kappa)}$ we sample $f_w(\kappa)$ far enough from the origin $w=0$ and for values of $w$ far enough from each other.

An exact relation (the statistical tilt symmetry~\cite{Schulz1988}) assures that the quadratic part of the Hamiltonian (and thus the constant $\kappa$) is not renormalized. This means that the length associated by a simple dimensional analysis  to the bare constant $\kappa$, namely $L_\kappa=c/\kappa$, corresponds to the correlation length of the system: above $L_\kappa$ the interface is flat and feels the harmonic parabola only, while below $L_\kappa$ the interface is rough with the characteristic roughness exponent at depinning $\zeta \simeq 0.39$~\cite{Rosso2002}.

\begin{figure}
\begin{center}
\includegraphics[width=.650\columnwidth]{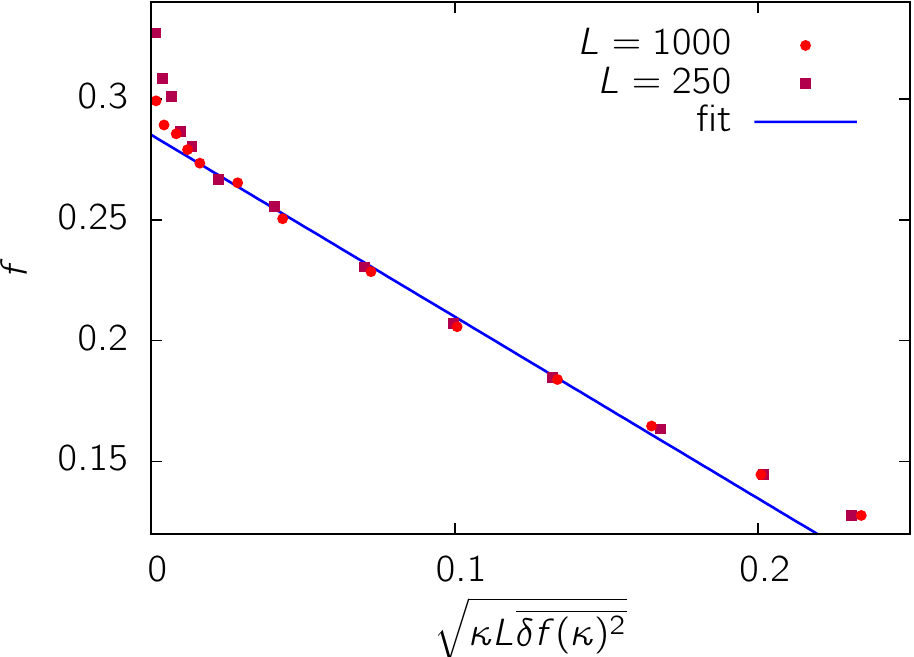}
\end{center}
\caption{(Colour online) Pinning force averaged over the parabola position $w$, versus a function of the spring stiffness. A fit of the linear part gives the critical force, see Eq.~\eref{eq:fkappafc}. This plot is for model A disorder with $\sigma=0.8$ and we found $f\ind{c}=0.285$.}
\label{fig_kappa_f}
\end{figure}

The separation from the critical depinning point (located exactly at the critical driving force $f\ind{c}$) is described by the power law scaling~\cite{Nattermann1992} $f-f\ind{c}\sim\xi^{-1/\nu}$ where~$\xi$ is the correlation length, given in our case by $L_\kappa$.
When $\kappa \to 0$ (while keeping $ L\gg L_\kappa$) the pinning force tends to the thermodynamical critical force $f\ind{c}$. Gathering the previous scalings, we thus have that the finite size effects on the force take the form:
\begin{eqnarray} \label{finitesize}
&& \overline{f(\kappa)}=f\ind{c} + c_1 \kappa^{1/\nu}+\cdots .
\end{eqnarray}
The fluctuations around this value,  $\overline{\delta f(\kappa)^2}$, depend on $L$ and $\kappa$.
In the limit $L\gg L_\kappa=1/\kappa$, the interface can be modeled as a collection of independent interfaces of size $L_\kappa$ and the central limit theorem assures that the variance $\overline{\delta f(\kappa)^2}$ should  scale as $\sim \kappa^{2/\nu}$, but  with an extra factor $L_\kappa/L$. This allows us to  write an extrapolation formula for~$f\ind{c}$ which is independent of the critical exponent $\nu$:
\begin{equation}
\overline{f(\kappa)}=f\ind{c}+c_1\sqrt{\kappa L  \overline{\delta f(\kappa)^2}}+\cdots
\label{eq:fkappafc}
\end{equation}
Our determination of $f\ind{c}$ is performed using this relation, by extrapolating the numerical measurements of $\overline{f(\kappa)}$ for different values of $\kappa$ to the limit $\kappa\to 0$, as shown on Fig.~\ref{fig_kappa_f}.
 It is worth noticing that most of the details of the finite size system such as
the boundary conditions or the presence of the parabolic well  does not affect the thermodynamic value of $f\ind{c}$ which depends only on the elastic constant $c$ and on the disorder statistics~\cite{Kolton2013, Budrikis2013}.
Our extrapolation of $f\ind{c}$, shown on Fig.~\ref{fig_kappa_f}, has been performed on samples of size $L=1000, 4000$ (depending on the value of sigma) and for parabola curvatures down to $\kappa=10^{-4}$.

\subsection{Results}

The dimensionless critical force $F\ind{c}=f\ind{c}/\sigma$ is plotted versus the dimensionless disorder parameter
\begin{equation}\label{eq:disorder_parameter}
\Sigma=\frac{\sigma\tilde\Delta_x(0)}{4\pi c}\int|k_u|\tilde\Delta_u(k_u)\dd k_u
\end{equation}
on Fig.~\ref{fig_sigma_fc}. In all three cases, the results are very close to the theoretical prediction $F\ind{c}=\Sigma$ (equivalent to~\eref{eq_anal_pred}) when the disorder parameter is small.

\section{Conclusion}\label{sec_conclu}

We have shown that the critical force for a long-range elastic line in a random landscape can be computed perturbatively in the collective pinning regime, yielding the expression (\ref{eq_anal_pred}). Our result for the critical force gives, together with its scaling with respect to the microscopic parameters, its dependence on the disorder geometry. 
Indeed, we have shown that two disorders that can be attributed the same correlation lengths (as in model A and~B) may present a different critical force that is precisely predicted by our theory.

Some previous studies have studied the scaling of the critical force with respect to microscopic parameters such as the disorder amplitude $\sigma$, the elastic constant $c$ and  disorder correlation lengths $\xi_x$ and $\xi_u$ in the directions $x$ and $u$~\cite{Larkin1979, Nattermann1990, Demery2012c}. In particular for the long-range elastic line, the following scaling has been found for the critical force in the collective pinning regime~\cite{Demery2012c}:
\begin{equation}\label{eq_fc_pheno}
f\ind{c}\sim \frac{\sigma^2\xi_x}{c\xi_u}.
\end{equation} 
The lengths $\xi_x$ and $\xi_u$  characterize the typical scale of the disorder correlation along $x$ and $u$, but these scales cannot be uniquely defined. 
Different definitions lead to correlation lengths that differ only by a numerical factor, so the scaling law (\ref{eq_fc_pheno}) holds independently of the chosen definitions. 
However, this prevents the use this scaling law to make a quantitative prediction. Our formula allows to overcome this problem.
 In particular, starting from Eq.~(\ref{eq_anal_pred}) and writing
\begin{equation}\label{eq_correl_adim}
\Delta_x(x) = \Delta_{x1}(x/\xi_x)
\end{equation}
where $\Delta_{x1}$ is a function of the dimensionless variable $x/\xi_x$ and a similar relation defines $\Delta_{u1}$, one gets
\begin{equation}
f\ind{c}= \left(\frac{\tilde\Delta_{x1}(0)}{4\pi}\int |q_u| \tilde\Delta_{u1}(q_u)\dd q_u\right) \times\frac{\sigma^2\xi_x}{c\xi_u}.
\end{equation}
This shows that our analytical prediction~(\ref{eq_anal_pred}) allows to recover the scaling law~(\ref{eq_fc_pheno}) and gives additionally the prefactor as a function of the correlation functions, \textit{i.e.}~it yields the explicit dependence of the critical force on the disorder geometry.

The present work is not the first attempt to compute the critical force perturbatively: expansions have been performed at weak disorder~\cite{Efetov1977,Chauve2000}, small temperature~\cite{Chen1995} or large velocity~\cite{Schmid1973,Larkin1974}. 
Weak disorder expansions are valid up to the Larkin length~\cite{Larkin1979}, $L_c$ defined as the distance at which the line wanders enough to see the finite disorder correlation length $\xi_u$, (namely $|u(L_c)-u(0)|\simeq \xi_u$). Above the Larkin length, they however predict an incorrect roughness exponent~\cite{Kardar1987}.
Last, large velocity expansions give an estimation of the critical force that is obtained by continuing a large-velocity asymptotic result, which lies very far from the depinning regime, to zero velocity.
Our computation does not need such continuation, and is compatible with the fact that perturbative expansions in the disorder amplitude are incorrect above the Larkin length, since the critical force can be evaluated from the line behaviour at the scale of the Larkin length~\cite{Larkin1979}.

Our analysis is a first step towards a more general understanding of the critical force dependence, and it can be extended in several directions. First, the opposite \textit{individual} pinning regime occurring at a high disorder amplitude is worth investigating. 
The perturbative analysis used here is not suited for its study, but a few comments can be made on the grounds of former numerical studies~\cite{Demery2012c, Patinet2013}. 
First, its scaling with respect to the disorder amplitude and correlation length has been elucidated for a long-range elastic line, giving~\cite{Demery2012c}
\begin{equation}
f\ind{c}\sim\sigma;
\end{equation}
thus the critical force is now proportional to the disorder amplitude and does not depend on the disorder correlation lengths. Moreover, we have shown in a previous study~\cite{Demery2012c} that the critical force is given by the strongest pinning sites if the pinning force is bounded. In this case, it is likely that the dependence on the disorder geometry is very weak. The case of unbounded pinning force requires further investigation.

Another issue arising from our study is the question of the landscape smoothness. Our analysis requires an expansion of the potential to the second order around the position of the unperturbed line (see Eq.~(\ref{eq_expansion})): the force generated by the potential must be continuous. 
When one tries to apply the analytical prediction~(\ref{eq_anal_pred}) to a discontinuous force landscape, it diverges because of a cusp present in the correlation function $\Delta_u(u)$. 
On the other hand, our previous numerical study~\cite{Demery2012c} used a discontinuous force landscape and did not reveal any divergence, while the dependence on the disorder amplitude, $f\ind{c}\sim\sigma^2$, was the same as the one observed here.
This suggests that the divergence obtained when we try to apply our result to a discontinuous force landscape is regularized by a mechanism that is out of reach of the present perturbative computation.
Understanding the behaviour of the elastic line and the critical force in a rougher force landscape would be an important advance on a theoretical point of view, but also for experiments where discontinuous force landscapes are ubiquitous~\cite{Soriano2002, Bonamy2008}. 

Last, the case of a short-range instead of long-range elasticity remains to be understood within our approach;
interesting comparison could be established with the characteristic force of the creep-regime, whose dependency in the details of the disorder correlator (for RB disorder) has been examined recently~\cite{agoritsas_temperature-induced_2010,agoritsas_static_2013,agoritsas_static_numeric_2013}.

\ack
We would like to thank E. Agoritsas for fruitful discussions and for a critical reading of the manuscript. V.~D. acknowledges support from the Institut des Systèmes Complexes de Paris Île de France.

\appendix

\section{Disorder generation and correlation}\label{ap_disorder_correlations}

We detail here the procedures used to generate the different models of disorder, and how to compute the disorder correlation function, that is needed to evaluate the critical force (\ref{eq_anal_pred}).

\subsection{Random field disorder: models A and B}

\begin{figure}
\begin{center}
\includegraphics[width=.750\columnwidth]{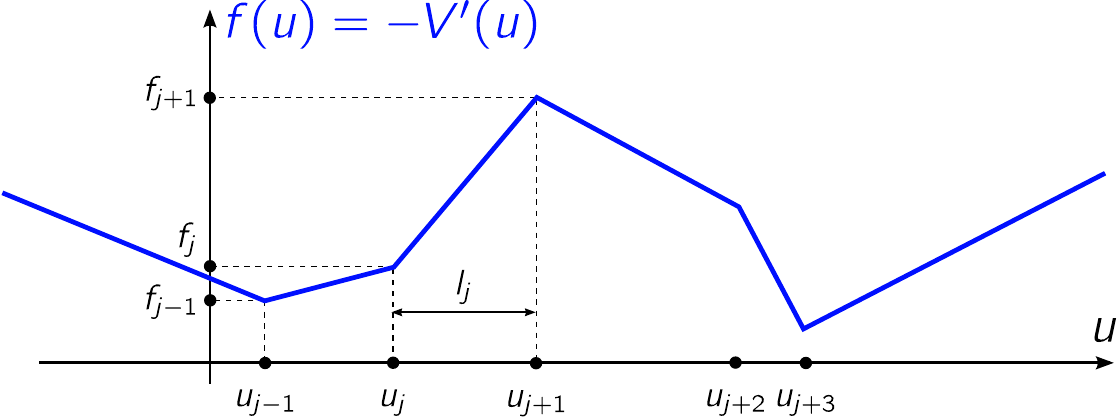}
\end{center}
\caption{(Colour online) Random force disorder on a rail: the force is continuous and piecewise linear.}
\label{fig_disorder_scheme}
\end{figure}

For a random force disorder, the disorder is generated on each rail using the following procedure (see Fig.~\ref{fig_disorder_scheme}):
\begin{itemize} \setlength{\itemsep}{1pt}
\item The rail is divided into segments of random length $l$ drawn in the distribution $P(l)$.
\item At the point linking the segment  $j-1$ and the segment  $j$, a random force $f_j$ is drawn from a Gaussian distribution with zero mean and unit variance.
\item Inside the segment $j$, at a generic point $u$, the force $f(u)$, is obtained by the linear interpolation such that $f(u_j)=f_j$ and $f(u_{j+1})=f_{j+1}$.
\end{itemize}

\begin{figure}
\begin{center}
\includegraphics[width=.650\columnwidth]{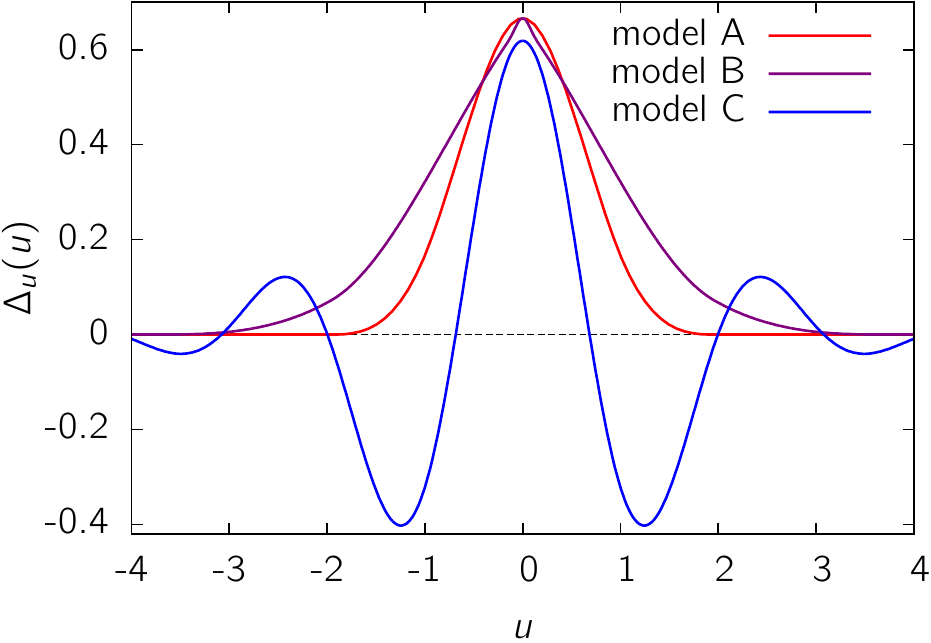}
\end{center}
\caption{(Colour online) Two-point correlation function of the force $\Delta_u(u)=\overline{\partial_u V(0)\partial_u V(u)}$ for the disorder models A, B and C. The model C correlation function has been rescaled by a factor $0.25$.}
\label{fig_correls}
\end{figure}

We want to know the correlation of the forces at two points separated by a distance $u\geq 0$, say $f(0)$ and $f(u)$. This correlation is non-zero if the two points lie on the same segment or on neighbour segments. We introduce the length $l$ of the segment where the point $0$ lies, the length $l'$ of its right neighbour, and the left end $u_0$ of the first segment. The probability distribution for $l$ is $Q(l)=lP(l)/\bar l$, where $\bar l=\int_0^\infty P(l)dl$; the probability distribution for $l'$ is simply $P(l')$ and the one of $u_0$ is $l^{-1}\chi_{[-l,0]}(u_0)$ (meaning that the point $0$ is uniformly distributed in its segment). Putting these probabilities together, we get the probability distribution for $(l,l',u_0)$:
\begin{equation}\label{eq_distrib_seglengths}
\mathbb{P}(l,l',u_0)=\bar l^{-1}P(l)P(l')\chi_{[-l,0]}(u_0).
\end{equation}

The points $0$ and $u$ are on the same segment if ${u\leq u_0+l}$. The force at $u_0$ is $f_0$ and the force at $u_0+l$ is $f_1$; $f_0$ and $f_1$ are uncorrelated random variables with zero mean and unit variance. The forces at $0$ and $u$ are
\begin{eqnarray}
f(0)&=f_0 \frac{u_0+l}{l}+f_1 \frac{-u_0}{l},\\
f(u)&=f_0 \frac{u_0+l-u}{l}+f_1 \frac{u-u_0}{l}.\\
\end{eqnarray}
The correlation between these two forces is
\begin{equation}\label{eq_correl_sameseg}
\overline{f(0)f(u)}=\frac{2u_0^2+2(l-u)u_0+l^2}{l^2}.
\end{equation}
Here, the average is restricted to the forces $f_0$ and $f_1$, the other variables $l$, $l'$ and $u_0$ are fixed.
On the other hand, when $u_0+l\leq u\leq u_0+l+l'$, the two points lie on neighbour segments. The same argument gives for the force correlation 
\begin{equation}\label{eq_correl_neighbourseg}
\overline{f(0)f(u)}=\frac{-u_0^2-(l+l'-u)u_0}{ll'}.
\end{equation}

Gathering the results (\ref{eq_distrib_seglengths},\ref{eq_correl_sameseg},\ref{eq_correl_neighbourseg}) and integrating over $u_0$ gives for the correlation function:
\begin{eqnarray}\label{eq_gen_correl_continuous force}
\hspace*{-25mm}
\Delta_u(u)=\frac{1}{\bar l}\int_0^\infty \dd l P(l)\left(\chi_{[0,l]}(u) \frac{u^3-3l^2u+2l^3}{3l^2}+
\vphantom{\left[\frac{-2u_0^3-3(l+l'-u)u_0^2}{6ll'} \right]^{\min(u-l,0)}_{\max(u-l-l',-l)}}
\right.
\nonumber\\
\hspace*{-2mm}
\left.
 \int_0^\infty \dd l'P(l')\chi_{[0,l+l']}(u) \left[\frac{-2u_0^3-3(l+l'-u)u_0^2}{6ll'} \right]^{\min(u-l,0)}_{\max(u-l-l',-l)} \right),
\end{eqnarray}
where we have used the notation $[g(u_0)]_a^b=g(b)-g(a)$.

For the model A, all the segments have the same length $l=1$, corresponding to the probability density
\begin{equation}\label{eq_distrib_A}
P(l)=\delta(l-1),
\end{equation}
For the model B, the segment lengths can take two values, $0.1$ and $1.9$, with probability $1/2$ each:  
\begin{equation}\label{eq_distrib_B}
P(l)=\frac{1}{2}\delta(l-0.1)+\frac{1}{2}\delta(l-1.9).
\end{equation}

For the model A, inserting the probability density (\ref{eq_distrib_A}) in the general formula (\ref{eq_gen_correl_continuous force}) gives the correlation function for $u\geq 0$,
\begin{equation}
\Delta_u(u)=\chi_{[0,1]}(u)\frac{3u^3-6u^2+4}{6}+\chi_{(1,2]}(u)\frac{(2-u)^3}{6},
\end{equation}
it is plotted on Fig.~\ref{fig_correls}.
To compute the critical force (\ref{eq_anal_pred}), we need the following quantity, 
\begin{eqnarray}
\int |k_u|\tilde\Delta_u(k_u)\dd k_u & =4\int_0^\infty \frac{\Delta_u(0)-\Delta_u(u)}{u^2}\dd u\nonumber\\
& =8\log(2).
\end{eqnarray}

For the model B, the correlation function is more complex and is plotted on Fig.~\ref{fig_correls}. The integral entering the expression~(\ref{eq_anal_pred}) of the critical force has to be computed numerically; we get
\begin{equation}
\int |k_u|\tilde\Delta_u(k_u)\dd k_u \simeq 6.91.
\end{equation}

\subsection{Random bond disorder: model C}

A random bond disorder can be generated on a rail by drawing random energies for points on a grid of step~$l$. A spline interpolation of these energies then allows to get a smooth landscape of potential. 
We determine here the two-point correlation function of such a disorder (see~\cite{agoritsas_static_numeric_2013} for a similar study for a two-dimensional spline).
Specifically, let us consider a grid of spacing $l$ with $2n+1$ points indexed from $-n$ to $n$.
A random value $V_i$ is attached to each site $u_i=il$ of the grid.
The function $V(u)$ is a cubic spline of the $(V_i)_{-n\leq i\leq n}$, that is:
\begin{itemize}\setlength{\itemsep}{1pt}
\item $V(u)$ is a cubic polynomial on each lattice segment $[u_i,u_{i+1}]$ for $-n\leq i<n$,
\item $V(u)$ is continuous on each lattice site $u_i$, and equal to $V_i$: $V(u_i^+)=V(u_i^-)=V_i$\,,
\item the first and second derivatives of $V(u)$ are continuous: $V'(u_i^+)=V'(u_i^-)$ and $V''(u_i^+)=V''(u_i^-)$.
\end{itemize}
One defines the coefficients $A_i^0,\ldots,A_i^3$ ($-n\leq i<n$) of the polynomials as
\begin{equation}
  \label{eq:defAiinterp}
   V(u)=A_i^0+A_i^1(u-u_i)+\frac{A_i^2}2(u-u_i)^2+\frac{A_i^3}{3!}(u-u_i)^3
\end{equation}
for $u_i\leq u<u_{i+1}$. One has $A_i^0=V_i$.

Denoting $l_i=u_{i+1}-u_i$ (not needed to be constant, we will keep it generic for a while), the continuity conditions write:
\begin{eqnarray}
  A_{i+1}^0&=A_i^0+l_iA_i^1+\frac 12l_i^2A_i^2+\frac 1{3!}l_i^3A_i^3 \:,
  \label{eq:splineA0}
\\
  A_{i+1}^1&=A_i^1+l_iA_i^2+\frac 12l_i^2A_i^3                      \:,
  \label{eq:splineA1}
\\
  A_{i+1}^2&=A_i^2+l_iA_i^3                                        \:.
  \label{eq:splineA2}
\end{eqnarray}

There are $6n$ unknown variables and $6n-2$ of those bulk equations.
They have to be complemented by boundary conditions (\textit{e.g.}~fixing the values of the derivatives
at extremities, or imposing periodic boundary conditions).
The simplest way to solve the set of equations is to eliminate the $A_i^1$'s and the  $A_i^3$'s
to obtain equations on the $A_i^2$'s only, as a function of the parameters $l_i$ and  $A_i^0=V_i$.
From~\eref{eq:splineA2} one has
$
  A_i^3=({A_{i+1}^2-A_i^2})/{l_i}
$
and substituting into~\eref{eq:splineA0} one obtains $A_i^1$:
\begin{equation}
  A_i^1=\frac{A_{i+1}^0-A_i^0}{l_i}-l_i\frac{2A_i^2+A_{i+1}^2}{6}.
  \label{eq:A2toA1}
\end{equation}
Using these expressions in~\eref{eq:splineA1} one gets the equations on the  $A_i^2$'s:
\begin{eqnarray}
  l_i A_i^2+2(l_{i}+l_{i+1})A_{i+1}^2+&l_{i+1}A_{i+2}^2=
 \nonumber\\
&6\frac{A_{i+2}^0-A_{i+1}^0}{l_{i+1}}-6\frac{A_{i+1}^0-A_{i}^0}{l_{i}}.
\end{eqnarray}
Those are quite complex to solve in general but simplifications occur for an uniform spacing $l_i=l$
and in the infinite grid size limit $n\to\infty$.

\textit{Solution for constant $l_i=l$}: The equations write
\begin{equation}
   A_i^2+4A_{i+1}^2+A_{i+2}^2=\frac 6{h^2}\big(A_{i}^0-2A_{i+1}^0+A_{i+2}^0\big).
\end{equation}
They take the form $M\vec {A^2}=\frac{6}{h^2}\Delta\vec {A^0} $ where $\Delta$ is the discrete Laplacian
and $M$ is a tridiagonal  $(2n+1)\times (2n+1)$ matrix. It is best represented as $M=6(\mathbf{1}+\frac{1}{6}\Delta) $ with
\begin{equation}
\Delta=
\left( \begin{array}{ccccc}
      -2&1&0&0&\ldots \\
      1&-2&1&0&\ldots \\
      0&1&-2&1&       \\
      \vdots&&\ddots&\ddots&\ddots
    \end{array}
\right)
\end{equation}
which allows to invert $M$ by writing:
\begin{equation}
  M^{-1}=\frac 16 \sum_{p\geq 0} \frac{(-1)^p}{6^p}\Delta^p .
\end{equation}
Hence, the vector $\vec {A^2}$ of the $A_i^2$'s is obtained as
\begin{equation}
  \vec {A^2}= \frac 1{l^2}\sum_{p\geq 0}\frac{(-1)^p}{6^p}\Delta^{p+1}  \vec {A^0}.
  \label{eq:A0A2}
\end{equation}

Each of the $A_i^2$'s is a linear combination of all the fixed potentials $A_i^0=V_i$'s.
It is known that the coefficients of $\Delta^p$ are given 
in the infinite size limit $n\to\infty$ by the binomial coefficients, up to a sign.
For instance the diagonal and subdiagonal elements are
\begin{equation}
  \big(\Delta^p)_{ii}= (-1)^p 
\binom{2p}{p}
\qquad 
  \big(\Delta^p)_{i,i+1}= (-1)^{p+1} \binom{2p}{p-1}\nonumber.
\end{equation}

One is now ready to determine the correlator of the potential.
On a generic interval $il\leq y\leq (i+1)l$ $(i>0)$ one has
\begin{eqnarray}
  V(u+\eta)=&\Big(i+1-\frac ul\Big)A_i^0-\Big(i-\frac ul\Big)A_{i+1}^0+\nonumber\\
  &\frac{(u-il)(u-(i+1)l)}{6l}\times\nonumber\\
  &\big[((2+i)l-u)A_i^2+(u-(i-1)l)A_{i+1}^2 \big]\,
\label{eq:VuAi02}
\end{eqnarray}
where $\eta$ is uniformly distributed on $[0,l]$ and allows to implement the statistical invariance by translation of the disorder (and generalizes the result of~\cite{agoritsas_static_numeric_2013}).
To determine the correlation function $\overline{V(u)V(u')}$, one thus has to identify the segments to which $u$ and $u'$ belong, and expanding~\eref{eq:VuAi02}, to determine averages of the form  $\overline{A_i^0A_j^2}$.
Those are obtained from the large-$n$ limit explicit form of~\eref{eq:A0A2}, which reads
\begin{eqnarray}
  A^2_j
  &= \ldots+(-1)^i\frac 1{l^2}\sum_{p\geq 0}\frac{1}{6^p}\binom{2p+2}{p-i+1} A^0_{j+i} + \ldots\nonumber\\
  &= \ldots+(-1)^{i+1}\frac {6\sqrt3}{l^2}\left(2-\sqrt3\right)^{i} A^0_{j+i} + \ldots
  \label{eq:resA2iA0iplusj}
\end{eqnarray}
which yields for instance $\overline{A_0^0A_i^2}= (-1)^{i+1}\frac{6\sqrt3}{l^2}\left(2-\sqrt3\right)^{i}$.
One obtains a cumbersome expression in real space, defined piecewise, that we do not reproduce here for clarity.
After Fourier transformation, the correlator $\tilde R_u$ is found to take a simple form
\begin{equation}
\tilde R_u(k_u)=\frac{9\sinc \left(\frac{k_u}{2} \right)^8}{(2+\cos(k_u))^2}\:,
\end{equation}
which we have checked numerically.
The force correlation function is shown on Fig.~\ref{fig_correls}; unlike the random field correlation functions, it presents negative parts indicating anticorrelations of the disorder (due to the spline continuity constraints).

\section*{References}

\bibliographystyle{plain_url}
\bibliography{biblio}

\end{document}